\documentclass[12pt]{iopart}
\usepackage{times}
\usepackage{graphicx}

\begin{document}

\title{Mixed-Symmetry Shell-Model Calculations}
\author{V. G. Gueorguiev and J. P. Draayer}
\address{Department of Physics and Astronomy, Louisiana State University,
\\ Baton Rouge, Louisiana 70803, USA}

\begin{abstract}
The one-dimensional harmonic oscillator in a box problem is used to
introduce the concept of an oblique-basis shell-model theory. The method
is applied to nuclei by combining traditional spherical shell-model
states with SU(3) collective configurations. An application to $^{24}$Mg,
using the realistic two-body interaction of Wildenthal, is used to
explore the validity of this oblique-basis, mixed-symmetry shell-model
concept. The applicability of the theory to the lower $pf$-shell nuclei
$^{44-48}$Ti and $^{48}$Cr using the Kuo-Brown-3 interaction is also
discussed. While these nuclei show strong SU(3) symmetry breaking due
mainly to the single-particle spin-orbit splitting, they continue to
yield enhanced B(E2) values not unlike those expected if the symmetry
were not broken. Other alternative basis sets are considered for future
oblique-basis shell-model calculations. The results suggest that an
oblique-basis, mixed-symmetry shell-model theory may prove to be useful
in situations where competing degrees of freedom dominate the dynamics.
\end{abstract}

Two dominate but often competing modes characterize the structure of
atomic nuclei. One is the single-particle shell structure underpinned by
the validity of the mean-field concept; the other is the many-particle
collective behavior manifested through nuclear deformation. The spherical
shell model is the theory of choice when single-particle behavior
dominates \cite{Whitehead-shell model}. When deformation dominates, the
Elliott SU(3) model can be used successfully \cite{Elliott's SU(3)
model}. This manifests itself in two dominant elements in the nuclear
Hamiltonian: the single-particle term, $H_{0}=\sum_{i}\varepsilon
_{i}n_{i}$, and a collective quadrupole-quadrupole interaction,
$H_{QQ}=Q\cdot Q$. It follows that a simplified Hamiltonian
$H=\sum_{i}\varepsilon _{i}n_{i}-\chi Q\cdot Q$ has two solvable limits
associated with these modes.

\begin{figure}[b]
\includegraphics[width=7.5cm,height=4.5cm]{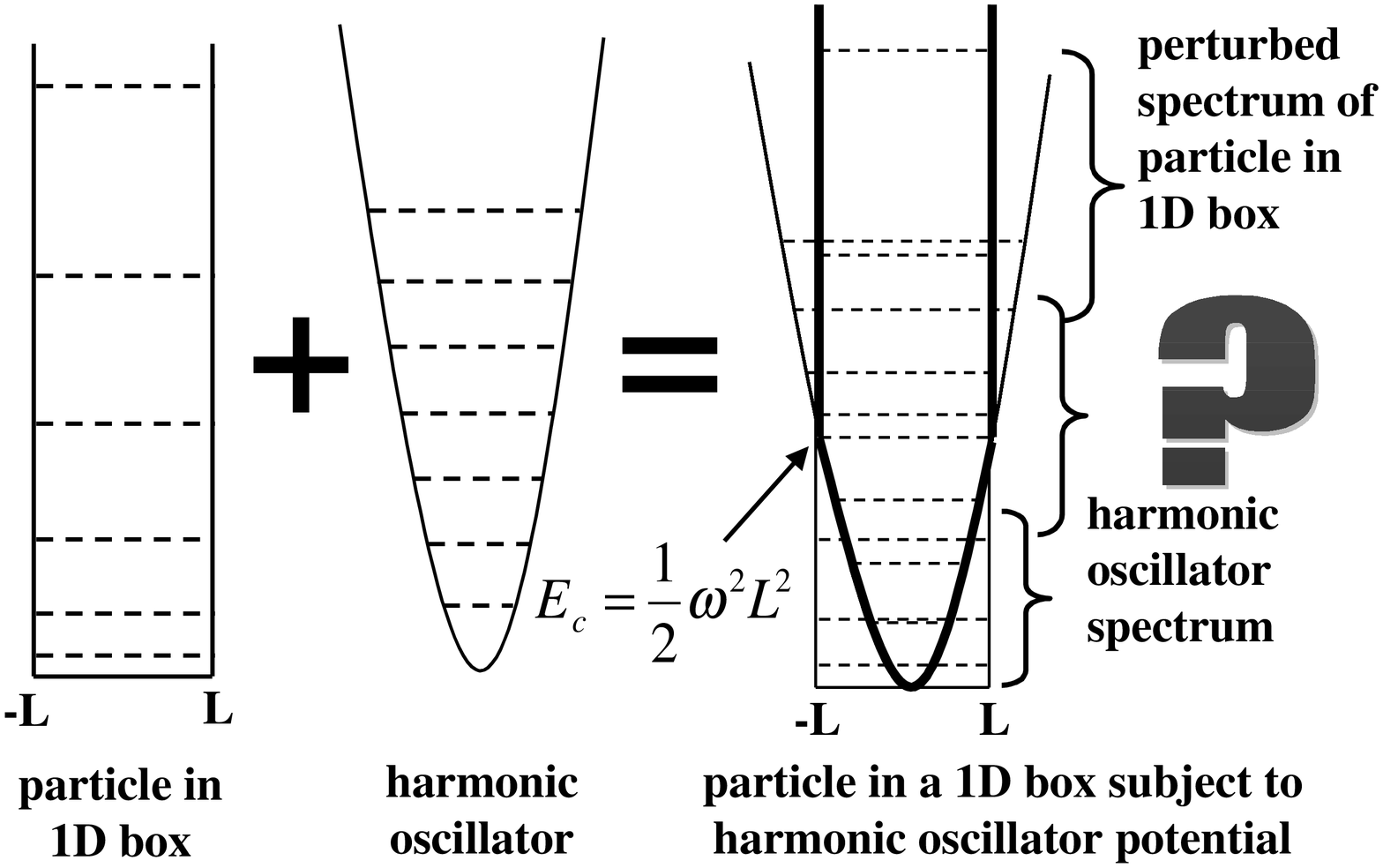}
\hfill
\includegraphics[width=7.5cm,height=4.5cm]{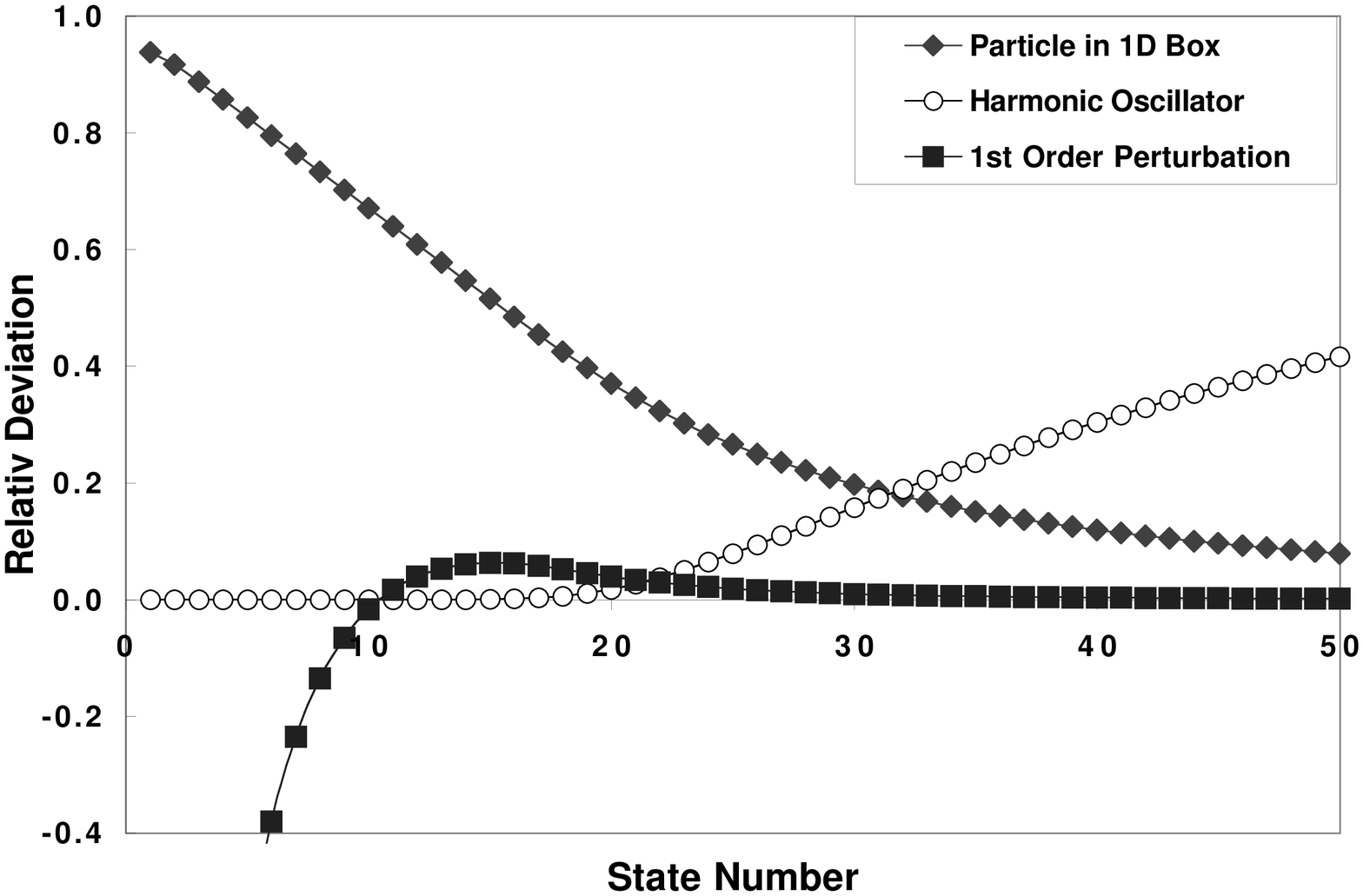}
\caption{Left graph shows the structure of the interaction potential of a
particle in an one-dimensional box subject to a harmonic oscillator
restoring force toward the center of the box. Right graph shows the
relative deviations from the exact energy eigenvalues for
$\omega =16$, $L=\pi /2$, $\hbar =m=1$. The open circles represent
deviation of the exact energy eigenvalue from the corresponding
harmonic-oscillator eigenvalue ($1-E_{ho}/E_{exact}$), the solid diamonds
are the corresponding relative deviation from the energy spectrum of a
particle in a 1D box, and the solid squares are the first-order
perturbation theory results.}
\label{HO1Dbox potential and spectrum}
\end{figure}

To probe the nature of such a system, we consider a simpler problem: the
one-dimensional harmonic oscillator in a box of size $2L$ \cite{Armen and
Rau}. As for real nuclei, this system has a finite volume and a restoring
force whose potential is of a harmonic oscillator type, $\omega ^{2}x^{2}/2$.
For this model, shown in fig.\ref{HO1Dbox potential and spectrum}, there
is a well-defined energy scale which measures the strength of the
potential at the boundary of the box, $E_{c}=\omega ^{2}L^{2}/2$. The
value of $E_{c}$ determines the type of low-energy excitations of the
system. Specifically, depending the value of $E_{c}$ there are three
spectral types:

\begin{itemize}
\item[(1)]  For $\omega \rightarrow 0$ the energy spectrum is simply that of
a particle in a box.

\item[(2)]  At some value of $\omega $, the energy spectrum begins with
$E_{c}$ followed by the spectrum of a particle in a box perturbed by the
harmonic oscillator potential.

\item[(3)]  For sufficiently large $\omega $ there is a harmonic oscillator
spectrum below $E_{c}$ followed by the perturbed spectrum of a 
particle in a box.
\end{itemize}

\begin{figure}[b]
\includegraphics[width=7.5cm,height=10cm]{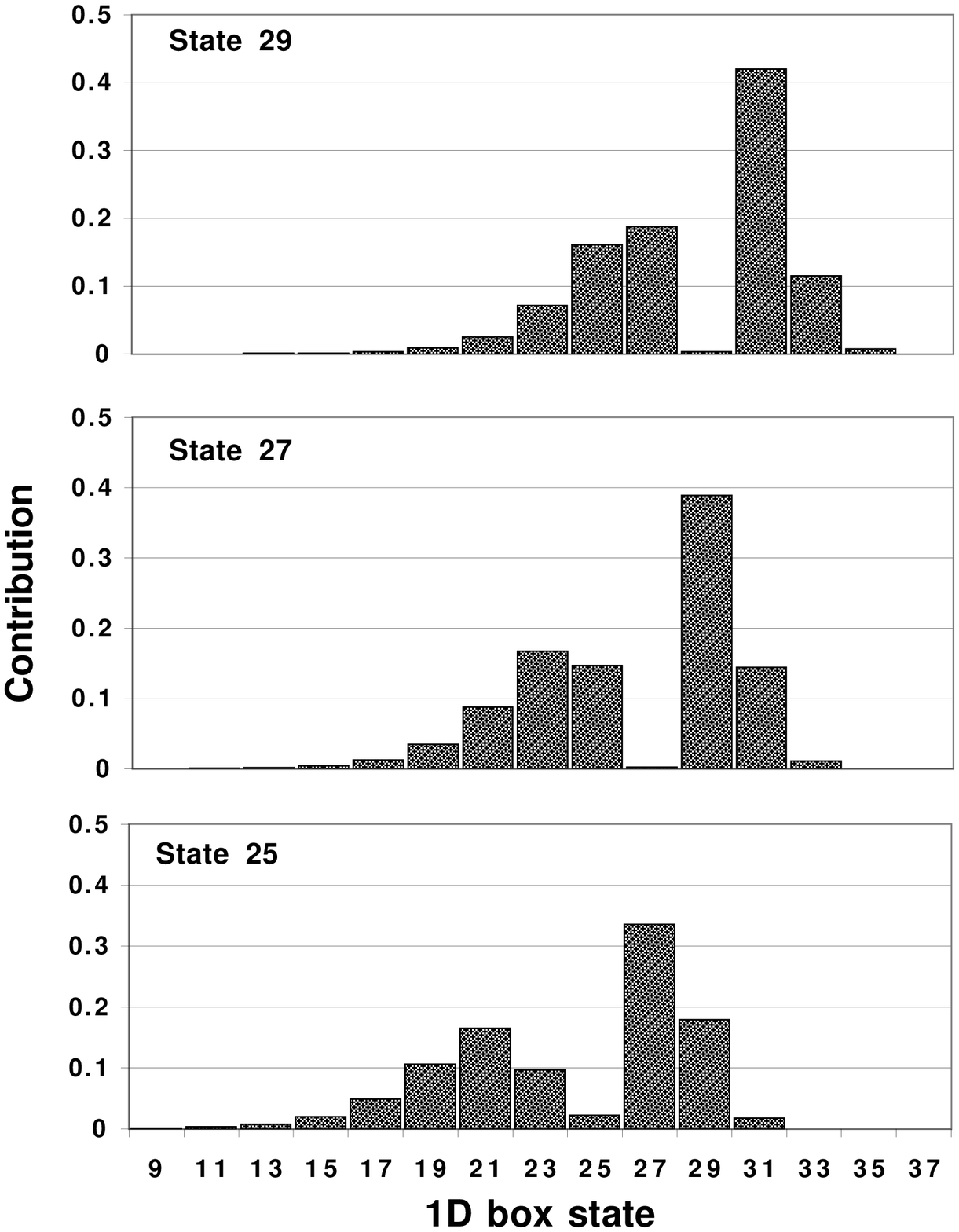}
\hfill
\includegraphics[width=7.5cm,height=10cm]{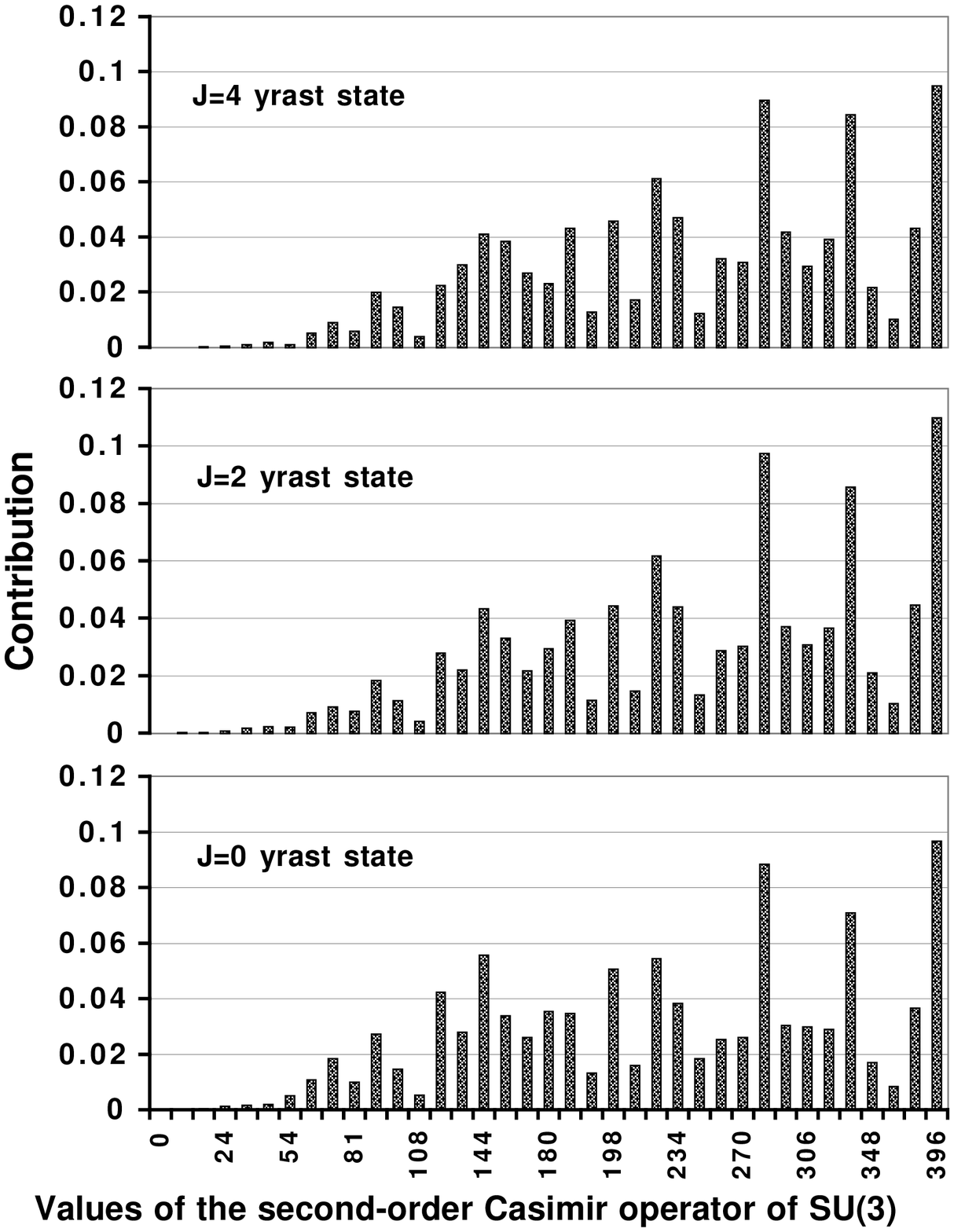}
\caption{Coherent structure with respect to the non-zero components of the
25th, 27th and 29th exact eigenvector in the basis of a free particle in
an one-dimensional box. Parameters of the toy Hamiltonian are
$\omega=16$, $L=\pi/2$, $\hbar=m=1$. Right graph shows
the coherent structure of the first three yrast states in $^{48}Cr$
calculated using realistic single-particle energies with Kuo-Brown-3 two
body interaction ($KB3$). On the horizontal axis is $C_{2}$ of $SU(3)$
with contribution of each $SU(3)$ state to the corresponding yrast state
on the vertical axis.}
\label{Coherent mixing}
\end{figure}

The last scenario (3) is the most interesting one since it provides an
example of a two-mode system. For this case the use of two sets of basis
vectors, one representing each of the two limits, has physical appeal,
especially at energies near $E_{c}.$ One basis set consists of the
harmonic oscillator states; the other set consists of basis states of a
particle in a box. We call this combination a mixed-mode / oblique-basis
approach. In general, the oblique-basis vectors form a nonorthogonal and
overcomplete set. Even thought a mixed spectrum is expected around
$E_{c}$, our numerical study, that includes up to 50 harmonic oscillator
states below $E_{c}$, shows that the first order perturbation theory in
energy using particle in a box wave functions as the zero order
approximation to the exact functions works quite well after the breakdown
of the harmonic oscillator spectrum. This observation is demonstrated in
the right graph of fig.\ref{HO1Dbox potential and spectrum} which shows
the relative deviations from the exact energy spectrum for a particle in a
box.

Although the spectrum seems to be well described using first order
perturbation theory based on particle in a box wave functions, the exact
wave functions near $E_{c}$ have an interesting structure. For example, the
zero order approximation to the wave function used to calculate the energy
may not be present at all in the structure of the exact wave function as
is the case shown in the left graph of fig.\ref{Coherent mixing}. Another
feature also seen in fig.\ref{Coherent mixing} is the common shape of the
distribution of the non-zero components along the particle in a box basis.
The right graph of fig.\ref{Coherent mixing} shows this same effect in
nuclei which is usually attributed to coherent mixing \cite{VGG
SU(3)andLSinPF-ShellNuclei,Adiabatic mixing}.

\begin{figure}[b]
\includegraphics[width=7.5cm,height=5cm]{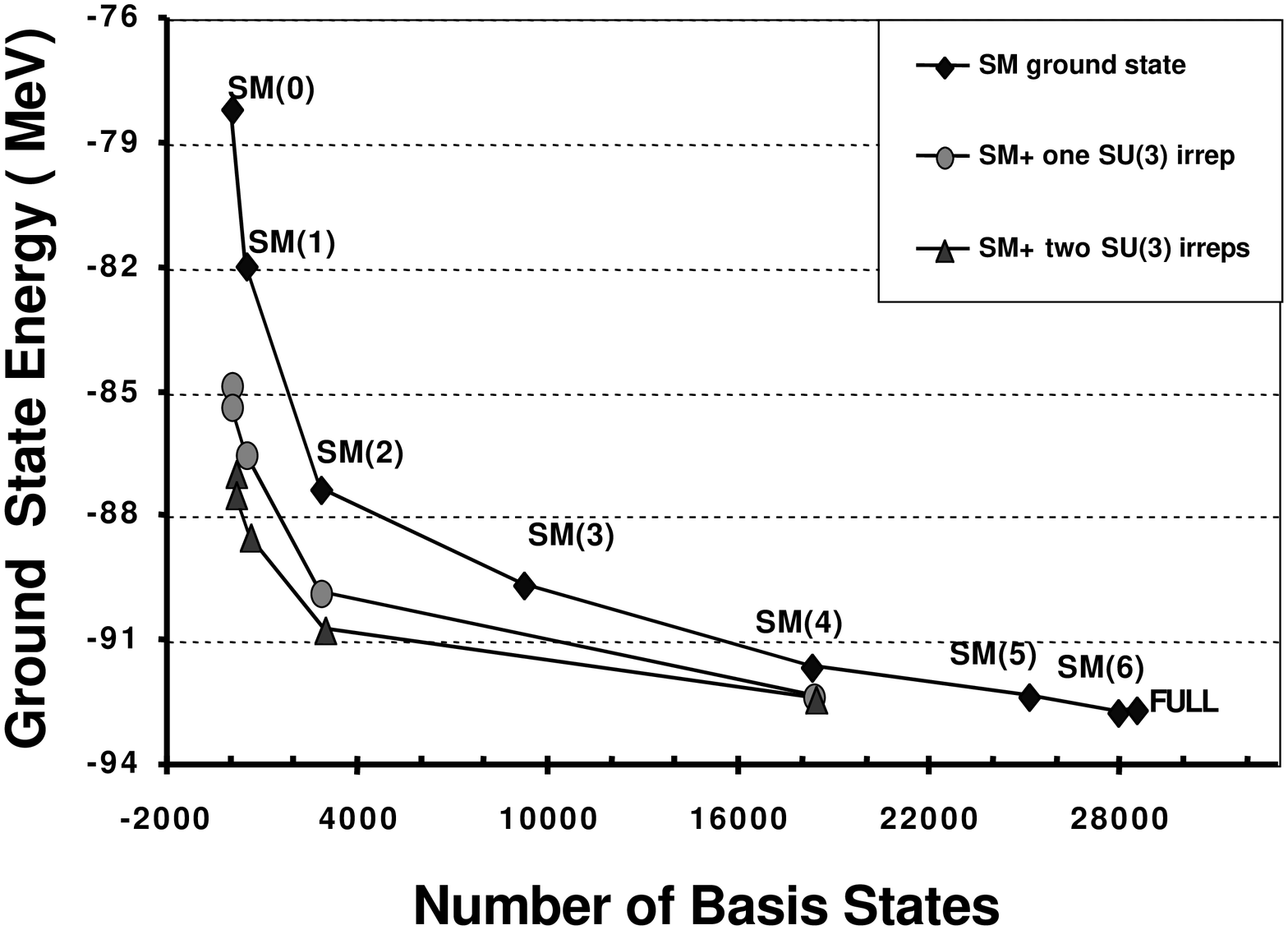}
\hfill
\includegraphics[width=7.5cm,height=5cm]{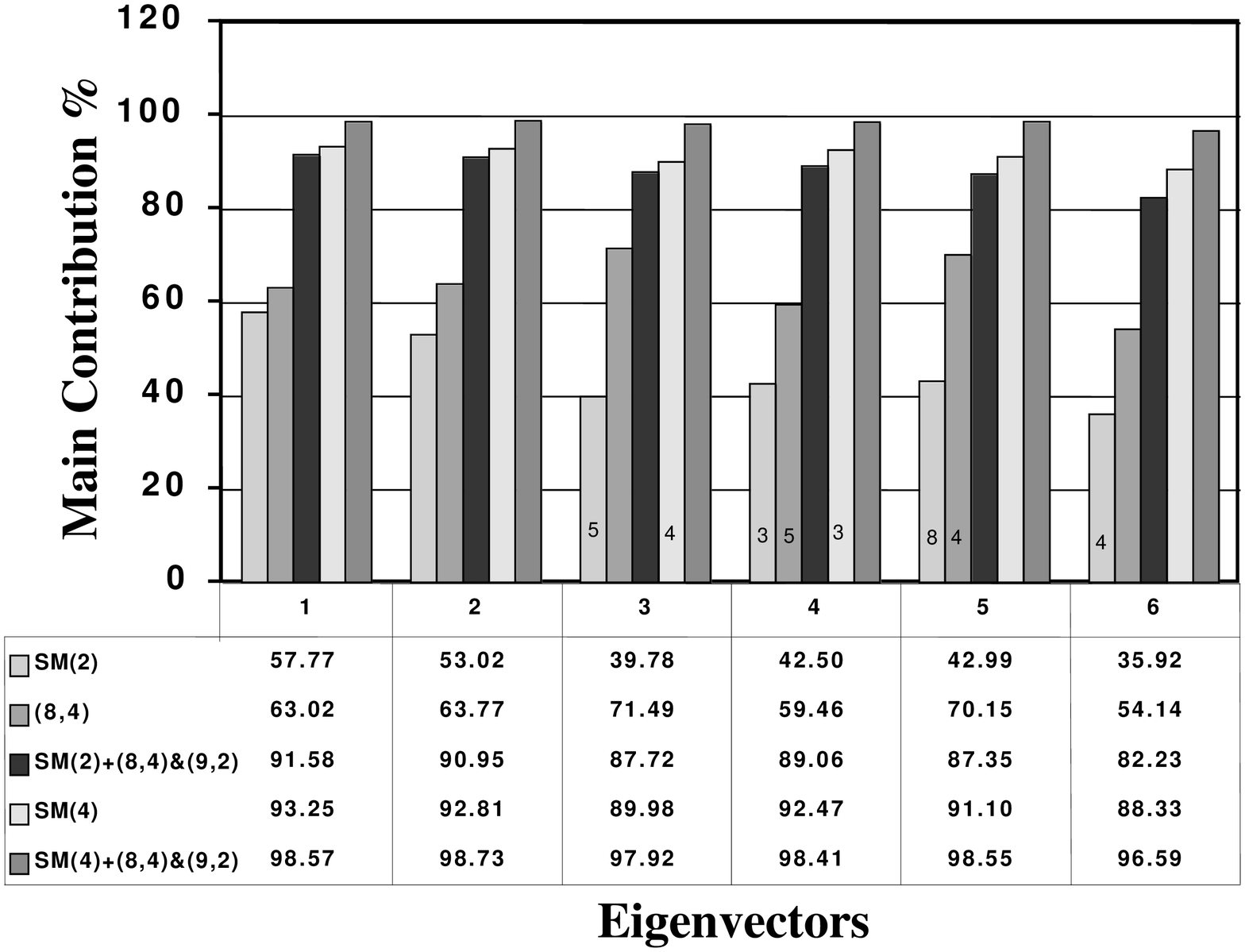}
\caption{
Left graph shows the calculated ground-state energy for $^{24}$Mg
as a function of various model spaces. SM(n) denotes spherical shell
model calculation with up to n particles outside of the $d_{5/2}$ sub-shell.
Note the dramatic increase in binding (3.3 MeV) in going from SM(2) to
SM(2)+(8,4)\&(9,2) (a 0.5\% increase in the dimensionality of the model
space). Enlarging the space from SM(2) to SM(4) (a 54\% increase in the
dimensionality of the model space) adds 4.2 MeV in the binding energy.
The right graph shows representative overlaps of pure
SM$(n)$, pure SU(3), and oblique-basis results with the exact full $sd$
shell eigenstates. A number within a bar denotes the state with the
overlap shown by the bar if it is different from the number for the exact
full-space calculation shown on the abscissa. For example, for SM(2) the
third eigenvector has the largest overlap with the fourth exact eigenstate,
not the third, while the fifth SM(2) eigenvector has greatest overlap
with the third exact eigenstate.}
\label{Mg24Oblique}
\end{figure}

An application of the theory to $^{24}$Mg \cite{VGG
24MgObliqueCalculations},  using the realistic two-body interaction of
Wildenthal \cite{Wildenthal}, demonstrates the validity of the mixed-mode
shell-model scheme. In this case the oblique-basis consists of the
traditional spherical states, which yield a diagonal representation of the
single-particle interaction, together with collective SU(3) configurations,
which yield a diagonal quadrupole-quadrupole interaction. The results shown
in fig.\ref{Mg24Oblique} were obtained in a space that
spans less than 10\% of the full-space.  They reproduce, within 2\% of the
full-space result, the correct binding energy as well as the low-energy
spectrum and the dominate structure of the states that have greater than
90\% overlap with the full-space results. In contrast, for a $m$-scheme
spherical shell-model calculation one needs about 60\% of the full space
to obtain results comparable with the oblique basis results.

Studies of the lower $pf$-shell nuclei $^{44-48}Ti$ and $^{48}Cr$ \cite{VGG
SU(3)andLSinPF-ShellNuclei}, using the realistic Kuo-Brown-3 (KB3)
interaction \cite{KB3 interaction}, show strong SU(3) symmetry breaking due
mainly to the single-particle spin-orbit splitting. Thus the KB3 Hamiltonian
could also be considered a two-mode system. This is further supported
by the behavior of the yrast band B(E2) values that seems to be insensitive
to fragmentation of the SU(3) symmetry. Specifically, the quadrupole
collectivity as measured by the B(E2) strengths remains high even though the
SU(3) symmetry is rather badly broken. This has been attributed to a
quasi-SU(3) symmetry \cite{Adiabatic mixing} where the observables behave
like a pure SU(3) symmetry while the true eigenvectors exhibit a strong
coherent structure with respect to each of the two bases. This provides
strong justification for further study of the implications of two-mode
shell-model studies.

Future research may provide justification for an extension of the theory to
multi-mode oblique shell-model calculations. An immediate extension of the
current scheme might use the eigenvectors of the pairing interaction
\cite{Dukelsky et al-Pairing} within the Sp(4) algebraic approach to the
nuclear structure \cite{Sviratcheva-sp(4)}, together with the collective
SU(3) states and spherical shell model states. Hamiltonian driven basis
sets can also be considered. In particular, the method may use eigenstates
of the very-near closed shell nuclei obtained from a full shell-model
calculation to form Hamiltonian driven J-pair states for mid-shell nuclei
\cite{Heyde's-shell model}. This type of extension would mimic the
Interacting Boson Model (IBM)
\cite{Iachello-1987} and the so-called broken-pair theory
\cite{Heyde's-shell model}. In particular, the three exact limits of the
IBM \cite {MoshinskyBookOnHO} can be considered to comprise a three-mode
system. Nonetheless, the real benefit of this approach is expected when
the system is far away of any exactly solvable limit of the Hamiltonian
and the spaces encountered are too large to allow for exact calculations.

\section*{Acknowledgments}

We acknowledge support from the U.S. National Science Foundation under Grant
No. PHY-9970769 and Cooperative Agreement No. EPS-9720652 that includes
matching from the Louisiana Board of Regents Support Fund. V. G. Gueorguiev
is grateful to the Louisiana State University Graduate School for awarding
him a dissertation fellowship and a travel grant to attend the XXIV
International Colloquium on Group Theoretical Methods in Physics held June
15-20, 2002 in Paris, France.

\end{document}